\documentstyle[aps,prl,multicol]{revtex}
\begin{document}

\title{Level statistics inside the core of a superconductive vortex}
\author{M.\ A.\ Skvortsov$^{1}$, V.\ E.\ Kravtsov$^{1,2}$ and M.\ V.\
Feigel'man$^{1}$}
\address{$^1$L. D. Landau Institute for Theoretical Physics, Moscow
117940, RUSSIA\\
$^2$International Center for Theoretical Physics, P.O. Box
586, 34100 Trieste, ITALY}
\date{\today}
\maketitle

\begin{abstract}
Microscopic theory of the type of Efetov's supermatrix sigma-model
is constructed for the low-lying electron states in
a mixed superconductive-normal system with disorder.
The developed technique is used for the study of the localized states
in the core of a vortex in a moderately clean superconductor
($1/\Delta \ll \tau \ll \omega_0^{-1} = E_F/\Delta^2$).
At sufficiently low energies $\epsilon \ll \omega_{Th}$,
the energy level statistics is described by the ``zero-dimensional"
limit of this supermatrix theory, with the effective ``Thouless energy"
$\omega_{Th} \sim (\omega_0/\tau)^{1/2}$.
Within this energy range the result for the density
of states is equivalent to that obtained within
Altland-Zirnbauer random matrix model of class C.
Nonzero modes of the sigma-model  increase the mean interlevel
distance $\omega_0$ by the relative amount of the order of
$[2\ln(1/\omega_0\tau)]^{-1}$.
\end{abstract}


There is a great deal of activity during the last 1.5 decades directed to the
study of electron energy levels and wavefunctions in disordered normal
metals \cite{efetov}, where they govern low-temperature transport properties.
In (s-wave) superconductors, disorder is usually of less importance,
since the excitation spectrum is gapful and single-electron states
are almost empty at $T \ll \Delta$. The situation is quite different
in mixed superconductive-normal systems (for recent reviews see
\cite{2}) where the gap in the excitation
spectrum can be: i) very low compared to the bulk $\Delta$, or
ii) just zero. The example of the first case is presented by an S-N-S sandwich
with the thickness $L_N$ of the N region much longer than the superconductive
 coherence length $\xi$ and the mean free path $l$.
Then at generic values  of the phase difference
$\varphi$ between superconductors the gap in the electron
spectrum in the N region
is of the order of the Thouless energy $E_{Th} = D/L_N^2 \ll \Delta$.
To calculate the density of states (DoS) fluctuations at $\epsilon > E_{Th}$,
and other mesoscopic effects in such systems, Altland, Simons and Taras-Semchuk
developed recently \cite{AST} a field theory which is a
version of Efetov's supersymmetric
matrix sigma-model explicitly taking into account the superconductive
coherence induced in the N metal due to the proximity effect. At low enough
energies the effective theory they derived is equivalent to
the Orthogonal random matrix ensemble at $\varphi=0$, and to the Unitary
ensemble at $\varphi \gg \delta/E_{Th}$ (where $\delta$ is the level spacing in the N).
Like in the usual normal-metal case \cite{efetov}, mesoscopic fluctuations
reveal themselves in the quantities containing products of {\it both}
retarded and advanced Green functions, whereas single, say, retarded
Green function is determined by the standard equations of the quasiclassical
theory of superconductivity.

Qualitatively different theoretical problem comes about in the second case,
ii), mentioned above, which is realized, e.g. in the same S-N-S sandwich at
$\varphi=\pi$ \cite{phi=pi}, or in a variety of situations where
the phase of the order parameter rotates due to the presence of an external
magnetic field.
Now the DoS is non-zero at arbitrary low energies, and quantum interference
due to Andreev scattering strongly
affects even the average DoS $\langle\rho(\epsilon)\rangle$ at
$\epsilon\sim\delta$. General approach to this kind of systems,
characterized by the zero {\it averaged over the whole system}
value of the superconductive
order parameter, was initiated by Altland and Zirnbauer (AZ) \cite{AZ},
who employed a generalized random-matrix approach. They have shown that
the particle-hole symmetry of the Bogolyubov-De Gennes (BdG) Hamiltonian
leads to important constraints  to be imposed on the random-matrix
Hamiltonians. Precise form of the  constraint depends on the presence
or absence of: a) time inversion symmetry, and b) spin rotation symmetry.
Thus AZ identified 4
additional (with respect to the standard Wigner-Dyson theory) classes
of random-matrix ensembles appropriate for the description of mesoscopic
fluctuations in this kind of S-N-S systems.
Crossover between such classes has been considered in \cite{FrB} using
the {\it space-independent} supermatrix sigma-model.
While the AZ approach is certainly highly suggestive, it has the same
general limitation as any ad hoc random matrix theory, i.e.\
the limits of its applicability to some real physical system are left
undetermined. Therefore we consider it highly desirable to develop
a fully microscopic field theory for the mesoscopic fluctuations in
S-N-S systems without a spectral gap.

In the present Letter we develop a microscopic
field-theory approach to the generic example of the ii) type of systems,
namely, to the core of a superconductive vortex.
It is known since Caroli, De Gennes and Matricon (CdGM) \cite{CdGM}
that the BdG equations for the electron states near the Abrikosov vortex
possess localized solutions with energies well below the bulk $\Delta$.
The spacing between these localized levels, $\omega_0$,
is of the order of $\Delta/(k_F\xi)$ and disappears in the quasiclassical
limit $k_F\xi \to \infty$. Thus it was tempting to consider the vortex core
as a kind of a ``normal tube" inside a superconductor \cite{BS},
and, indeed,
in many cases such a simplified picture was found \cite{KK} to be
at least qualitatively correct. Later it was demonstrated \cite{volovik}
that the presence of a quasi-continuum spectrum branch localized
on the vortex follows from general topological arguments; actually,
the number of such ``chiral" branches coincides with the topological
charge of the vortex.
However, it is not always possible to consider the chiral branch as a
continuos one, as it stands in the quasiclassical calculations.
It was shown recently
\cite{FeiSkv,Lnew}, that the discreteness of the localized
energy levels becomes of real importance in layered superconductors
at sufficiently low temperatures. In the previous paper \cite{FeiSkv}
we employed the AZ phenomenological approach to find
low-current nonlinearities in the current-voltage relation in a mixed
state of a moderately
clean superconductor  (the mean free path $l \gg \xi$, but
$l \ll \xi (k_F\xi)$). In such a case the inverse elastic
 scattering time $1/\tau$ is much larger than interlevel spacing $\omega_0$,
therefore the applicability of an appropriate random-matrix model
(which is, in fact, class C of the AZ classification) seems quite natural.
Vortex-induced dissipation in another limiting case of a
super-clean superconductor with extremely low concentration of impurities
($l \gg k_F\xi^2$) was considered recently by Larkin
and co-workers\cite{Lnew}. Here the system of electron levels in the
core was found to be extremely strongly correlated and almost integrable,
with the properties qualitatively different from the moderately clean
case \cite{FeiSkv}.
In the present Letter we again consider a
moderately clean limit $\omega_0 \ll 1/\tau \ll \Delta$, now within
microscopic approach starting from the BdG equations in the presence of
impurities-induced Gaussian random potential.
We derive the conditions
under which the AZ class C statistics is indeed realized in the vortex
core, and estimate the scale of non-universal corrections to it.
Throughout all the paper we consider purely 2-dimensional superconductor,
which is a good approximation for the case of sufficiently strong
layered anisotropy; more quantitative conditions can be found in
\cite{FeiSkv}.

Below we present a brief description of our method and results,
whereas their detailed presentation
is postponed to a forecoming paper \cite{full}.
It was already mentioned above that in the present problem even the
calculation of the average single-particle quantities is not trivial
and cannot be done within the quasiclassical theory, as long as low energies
comparable to the level spacing $\omega_0$ are considered.
Thus our goal here is to derive a field-theory technique
for the calculation of the average DoS
$\langle\rho(\epsilon)\rangle
= \langle\sum_j\delta(\epsilon-\epsilon_j)\rangle$.
To average the Green function over disorder, we use a standard
trick \cite{efetov} of representing it as the functional integral
over both Grassmann ($\chi$) and usual complex ($S$) fields which combine
into the superfield $\Phi$. The most direct way would be to work with
real-space-dependent superfield $\Phi({\bf r})$ corresponding to the
standard representation of the BdG Hamiltonian.
In this way we would obtain a field theory in terms of
 $Q({\bf r})$ supermatrix, depending on two spatial coordinates $r_x,r_y$.
On the other hand, low-lying states of the chiral branch depend upon
a single quantum number only (in the absence of disorder it is
just the angular momentum), as well as for a generic 1D problem.
Therefore, in the basis of such states the BdG Hamiltonian can be
represented as a random $N\times N$ Hermitean matrix
(where $N\sim \Delta/\omega_0$ is the total number of the localized
states in the core)
of the certain structure and
symmetry which we will discuss below. In the clean limit, $1/\tau \ll
\Delta$, the admixture of delocalized
($\epsilon > \Delta$) states to the low-lying ones can be neglected.
Thus it is convenient first to reduce the full 2D problem to
a sort of random matrix problem that can be further reduced to
the 1D field theory, explicitly containing the chiral spectrum branch
only.

In the basis of the chiral CdGM states
$\Psi_{\mu}({\bf r}) =
A (J_{\mu-1/2}(k_Fr), J_{\mu+1/2}(k_Fr))^T e^{i\mu\theta} e^{-K(r)}$
determined in \cite{CdGM} (here $A\sim\sqrt{k_F/\xi}$ is the normalization
constant, $\theta$ is the azimuthal angle in the real space, $\mu\in
[-N/2,N/2]$ is the angular momentum that takes half-integer values,
and $K(r)=(1/\hbar v_F)\int_0^r\Delta(r') dr'$),
the full Hamiltonian takes the form
$\langle\mu|\hat{H}|\mu'\rangle =
\omega_0\mu\delta_{\mu,\mu'} + \langle\mu|\hat{V}|\mu'\rangle$
where the second term is due to the random white-noise impurity potential
$U({\bf r})$ with the variance
$\langle U({\bf r})U({\bf r}')\rangle =
\delta({\bf r}-{\bf r}')/(2\pi\nu\tau)$;
correspondingly, in the functional integral one should use
$\mu$-dependent supervector $\Phi_{\mu}$
instead of the superfield $\Phi({\bf r})$.
This Hamiltonian obeys the symmetry
\begin{equation}
  \hat{H}=-\hat{\gamma}\hat{H}^T\hat{\gamma}^T; \quad
  \langle\mu|\hat{\gamma}|\mu'\rangle =(-1)^{\mu+1/2}\delta_{\mu+\mu'} ,
\label{gamma}
\end{equation}
which follows from an identity $\Psi_{-\mu}({\bf r})
=(-1)^{\mu+\frac{1}{2}}i\tau_{y}\Psi_{\mu}^{*}({\bf r})$ that reflects
the basic symmetry property of the BdG
Hamiltonian. Here we introduced the Pauli matrices $\tau_{x,y,z}$
in the $2\times 2$ Nambu space.

The standard way to solve a complicated random matrix problem is to
represent it in a form of the effective field theory.
In order to reduce the random matrix problem given by Eq.~(\ref{gamma})
to the 1D field theory
we make a continuous Fourier transform (considering $N$ as very large,
which is possible since the energy range $\epsilon \ll \Delta$ is studied
below) from the momentum variable $\mu$ to the ``angle" $\phi\in
[0,2\pi)$,
so our superfield will be defined as
$\Phi (\phi)=\sum_{\mu}\Phi_{\mu}e^{-i(\mu-1/2)\phi}$. Now we can write
down
an expression for the `partition function'
($\epsilon_+\equiv\epsilon+i\delta$):
\begin{equation}
  Z^R(\epsilon) = \int \exp i
    \left\{ \int \frac{d\phi}{2\pi}
      \Phi^*(\phi)
        \left(
          \epsilon_+ - i\omega_0 \frac{\partial}{\partial \phi} -
          \frac{\omega_0}2
        \right)
      \Phi(\phi)
    - \int\!\!\int \frac{d\phi d\phi'}{(2\pi)^2}
      \Phi^*(\phi) V(\phi,\phi') \Phi(\phi')
    \right\} D\Phi^*(\phi) D\Phi(\phi) .
\label{Z}
\end{equation}

Matrix elements $V(\phi,\phi')$ of the random potential in the
$\phi$-space obey
the symmetry relationship that follows from Eq.~(\ref{gamma}) and  are
given by
\begin{equation}
  V(\phi,\phi') = - e^{i(\phi-\phi')}V^{*}(\phi+\pi,\phi'+\pi)= A^2\int
d^2{\bf r}
w_{\phi \phi'}(r,\theta) U({\bf r}) e^{-2K(r)}.
\label{Vxx}
\end{equation}
The function $w_{\phi \phi'}$ can be presented, using summation formulae
for
the Bessel functions, as
\begin{equation}
\label{w-final}
  w_{\phi \phi'}(r,\theta) = \bigl( 1-e^{i(\phi-\phi')} \bigr) \:
  \exp \, \Bigl\{ -ik_Fr
    \bigl[
      (\sin \phi - \sin \phi') \cos\theta +
      (\cos \phi - \cos \phi') \sin\theta
    \bigr]
  \Bigr\}.
\label{w}
\end{equation}
All features of the theory are encoded in the pair correlator
${\cal W}(\phi_1, \phi_2, \phi_3, \phi_4) =
\langle V(\phi_1, \phi_2) V(\phi_3, \phi_4) \rangle$
where the averaging is performed
over the Gaussian distribution of the impurity's
potential $U({\bf r})$.
Since the typical value of $k_{F}r\sim k_{F}\xi\gg 1$, the correlator
${\cal W}(\phi_1, \phi_2, \phi_3, \phi_4)$ is essentially non-zero only
when the
oscillating exponents in
Eq. (\ref{w}) nearly cancel each other, i.e.\
when its arguments $\phi_{i}$ are nearly pair-wise coinciding \cite{full}:
\begin{equation}
\label{W}
  {\cal W}(\phi_1, \phi_2, \phi_3, \phi_4) = \frac{g\omega_0^2}{\pi}
T(\phi_1-\phi_2+\pi)
  (2\pi)^2
  \left[
    \delta(\phi_1-\phi_4) \delta(\phi_2-\phi_3)
    - e^{i(\phi_1-\phi_2)} \delta(\phi_1-\phi_3+\pi)
\delta(\phi_2-\phi_4+\pi)
  \right] ,
\end{equation}
where $g = 2A^4/\pi\nu\tau\omega_0^2k_F^2\sim 1/\omega_0\tau \gg 1$
and the kernel $T$ is given by
\begin{equation}
\label{kernel}
  T(\phi) =
  \cases{
    \frac{\pi}{2} \left|\cot\frac{\phi}2\right|,
      & if $|\phi|>\frac1{\sqrt{N}}$; \cr
    \sim \sqrt{N},
      & if $|\phi|<\frac1{\sqrt{N}}$. \cr
  }
\end{equation}
The $\delta$-function approximation (\ref{W}) for the correlator
${\cal W}$ is valid as long as the scale of the angular
variations of the field $\Phi(\phi)$ (below it will be seen to be
$\ell=[g\ln(N/g)]^{-1}$) is longer than the actual\cite{full}
 width $w(\phi_{1}-\phi_{2})\sim |N \sin(\phi_{1}-\phi_{2})|^{-1}$ of
those
$\delta$-functions.
Thus the following derivation is strictly valid under the condition
$w(\ell)\ll \ell$ which is equivalent to:
\begin{equation}
  \tau\sqrt{\omega_0\Delta} \gg \ln\Delta\tau.
\label{condition}
\end{equation}
Note that this condition is stronger than just the clean limit
condition $\tau\Delta \gg 1$. In what follows we will assume this
condition to be always fulfilled.

The next step of the $\sigma$-model derivation is to average the partition
function (\ref{Z}) using Eqs.~(\ref{W}), (\ref{kernel}). Before doing that
we need to take explicitly into account the symmetry (\ref{gamma}),
(\ref{Vxx})
which amount to the doubling of the number of components of the supervector
$\Phi(\phi)$ (a similar procedure in the standard approach \cite{efetov}
is related with the time-reversal symmetry).
Thus we introduce (cf.~with a similar procedure in \cite{FrB})
an additional $2\times 2$ ``particle-hole" (PH) space and
define a 4-dimensional supervector
field $\psi(\phi) = 2^{-1/2}\bigl(\Phi(\phi),
e^{i\phi}\Phi^*(\phi+\pi)\bigr)^T$.
Next we define the bar-conjugated superfield as
$\bar{\psi}(\phi) =\psi^+\sigma_{z}=[C(\phi)\psi(\phi+\pi)]^T$
with $C(\phi)=-e^{-i\phi}\sigma_zC_0$, where $\sigma_z$ is the Pauli matrix
acting in the PH space, and
\begin{equation}
\label{C0}
  C_0 = \left( \begin{array}{cc} 0 & 1 \\ k & 0 \end{array} \right)_{ph} ,
  \qquad
  k = \left(\begin{array}{cc} 1 & 0 \\ 0 & -1 \end{array}\right)_{fb} ,
\end{equation}
with FB meaning the Fermi-Bose space.
After the averaging over disorder the effective action
${\cal A}\{\psi\}$ entering the functional integral
for the retarded Green function
${\cal G}^R(\epsilon) = -i\int \Phi_a \Phi^+_a
\exp[\cal{A}\{\psi\}] D\psi^* D\psi$
(where $a$ implies either bosonic or fermionic component)
can be written as
\begin{equation}
  {\cal A}\{\psi\} = i
  \int \frac{d\phi}{2\pi}
    \bar\psi(\phi)
      \left(
        \epsilon_+ \sigma_z
        - i\omega_0 \frac{\partial}{\partial \phi} - \frac{\omega_0}{2}
      \right)
    \psi(\phi) -
   \frac{g \omega_{0}^2}{\pi} \int\!\!\int \frac{d\phi_1
d\phi_2}{(2\pi)^2}
    T(\phi_1-\phi_2+\pi) \bar\psi(\phi_1) \psi(\phi_2) \bar\psi(\phi_2)
\psi(\phi_1) .
\label{Action}
\end{equation}
The second term in the action (\ref{Action}) is very similar to that of
the 1D tight-binding model with off-diagonal random matrix elements
with variance decaying as $1/|x|$, as long
as we are interested in the scale $|x|\equiv |\phi_{1}-\phi_{2}+\pi|\ll
\pi$.
Therefore, we expect the usual 1D
localization to be absent in our problem just because of the long-range
nature of the off-diagonal disorder (cf.~\cite{MirFed}).

There is also another suggestive way of considering this term, which
helps to gain some intuition about its effect.
Namely, one can think
of the variable $\phi$ as an angle associated with the 2D quasiparticle
momentum $p=k_{F}\{\cos\phi, \sin\phi \}$.
Then the second term in Eq.~(\ref{Action})
corresponds to a 2D {\it particle-hole} scattering which is strongly enchanced
in the forward direction. For such a singular scattering one has to define
two scattering lengths $\ell$ and $\ell_{tr}\gg \ell$
(cf.\ with a similar situation discussed in \cite{AMW}):
$1/\ell \propto g \int d\phi\; \sigma(\phi) = g \ln(N/g)$, and
$1/\ell_{tr}\propto g\int d\phi\; \sigma(\phi)\; (1-\cos\phi)= g\gg 1$,
where $\sigma(\phi)$ is the differential cross-section and
$\phi=\phi_{1}-\phi_{2}+\pi$. For the careful evaluation of the
logarithmically divergent scattering rate $1/\ell$, one should use the
self-consistent Born approximation (SCBA) which takes into account both
terms in Eq.~(\ref{W}). It is equivalent to taking into account of the
``non-crossing" diagrams that can be generated by a
perturbative expansion of $\exp[\cal{A}\{\psi\}]$ in powers of $g$.
For $\epsilon/\omega_{0}\ll 1/\ell$, the ``crossing diagrams" of the same
order in $g$ turn out to be small by
the parameter $\ell/\ell_{tr}=1/\ln(N/g)$. It stands for the usual
quasiclassical parameter $(k_{F}\ell_{tr})^{1-d}$ in this effectively 1D
problem.

The existence of the small papameter
$1/\ln(N/g)=1/\ln\Delta\tau$ that allows to neglect the
``crossing diagrams", implies that one can derive an effective field theory
(nonlinear sigma-model) which describes the low-energy behavior of the
averaged Green function $G^{R}(\epsilon)$ for
$\epsilon/\omega_{0}\ll 1/\ell=g\ln(N/g)$.
This can be done in a standard way \cite{efetov} by the
Hubbard-Stratonovich decoupling of the quartic
term in Eq.~(\ref{Action}) and a further saddle point approximation
controlled by the parameter $1/\ln(N/g)$. Because of the symmetry
relationship (\ref{gamma}), (\ref{Vxx}) and the corresponding
relationship between $\bar{\psi}$ and $\psi$, one has to
perform both the local decoupling
containing $P(\phi)\;\psi(\phi)\otimes\bar{\psi}(\phi)$ and the non-local
one containing
$R(\phi_1,\phi_2)\;\psi(\phi_{1})\otimes\psi^T(\phi_{2})$.
Under the condition $\ell \gg 1/\sqrt{N}$ given by Eq.~(\ref{condition}),
both decouplings are important in order to obtain a correct form for the
imaginary part of the Green function in a saddle-point approximation
$P(\phi)=P_{0}$, $R(\phi_1,\phi_2)=R_{0}(\phi_1-\phi_2)$ which
is equivalent to the SCBA:
\begin{equation}
  G_{\epsilon}(\phi) = - \frac{2\pi i}{\omega_0} \sigma_z \theta(-\sigma_z\phi)
    e^{-|\phi|/\ell} e^{-i\frac{\epsilon}{\omega_0}\phi}, \quad
  P_0 = \frac{T_0}{\omega_0} \sigma_z, \quad
  R_0(\phi) = \frac{i}{\pi} T(\phi) G_{\epsilon}(-\phi) ,
\label{spsol}
\end{equation}
where $\ell^{-1}=g\ln(N/g)$, and $T_0 =
\int_0^{2\pi}T(\phi)\frac{d\phi}{2\pi} \approx
 \frac{1}{2}\ln N$.
In general, $m$-th Fourier harmonics of the kernel $T(\phi)$ is given by
$T_m \approx \ln\frac{\sqrt{N}}{|m|}$ for $1 \ll m \ll \sqrt{N}$.

Mesoscopic fluctuations are known\cite{efetov} to be described by
the slow rotations of the saddle-point solution, which are represented
in our case as
$  P(\phi) = U^{-1}(\phi) P_0 U(\phi), \quad
  R(\phi,\phi') = U^{-1}(\phi) R_0(\phi-\phi') U(\phi') $.
The corresponding action that describes the low-energy spectral properties
of the CdGM levels, reads:
\begin{equation}
\label{sigma-model}
  {\cal A}_\sigma[Q,U] = - \frac{\pi g}4 T_0^2
    \int\!\!\int \frac{d\phi_1d\phi_2}{(2\pi)^2}
    T^{-1}(\phi_1-\phi_2+\pi) \,{\rm Str}\, Q(\phi_1) Q(\phi_2)
  - \frac{\pi i}{2}
  \int \frac{d\phi}{2\pi} \,{\rm Str}
    \left(
      \frac{\epsilon}{\omega_0} \sigma_z Q(\phi)
      - i \sigma_z U(\phi) \frac{\partial U^{-1}(\phi)}{\partial \phi}
    \right),
\end{equation}
where
$Q(\phi) = U^{-1}(\phi)\sigma_z U(\phi)$,
and $U(\phi)$ is a $\pi$-periodic,
pseudo-unitary ($U^{-1}(\phi)=\bar{U}(\phi)$) matrix.
The action (\ref{sigma-model}) is valid for the energies
$\epsilon \ll \omega_0 /\ell = \tau^{-1}\ln\Delta\tau$.

The supermatrix $Q$ can be represented in the form
$Q(\phi) =\sigma_z[1+W(\phi)+\frac{1}{2}W^2(\phi) + O(W^3)]$ with
the supermatrix $W$ being purely off-diagonal in the PH space.
Then the symmetry $Q=\bar{Q}$ and convergence arguments lead to the
following form for the $W_{ph}$ and $W_{hp}$ blocks:
\begin{equation}
\label{W-param}
  W_{ph}(\phi) =
    \left(
      \begin{array}{cc} iz(\phi) & \alpha_1(\phi) \\ \alpha_1(\phi) & 0
      \end{array}
    \right)_{fb} , \qquad
  W_{hp}(\phi) =
    \left(
      \begin{array}{cc} iz^*(\phi) & \alpha_2(\phi) \\ -\alpha_2(\phi) & 0
      \end{array}
    \right)_{fb} .
\end{equation}
Here $z$ is a complex number and $\alpha_i$ are Grassmann numbers.
Expanding over $W(\phi)$, we obtain in the quadratic approximation
\begin{equation}
  {\cal A}_2[W_m] = \frac\pi4
    \,{\rm Str} \sum_m
      \left\{
        2g \left( \sum_{k=0}^{|m|-1} \frac{1}{2k+1} \right)
        + i \left( m \sigma_z - \frac{\epsilon}{\omega_0} \right)
      \right\}
    W_{2m} W_{-2m} ,
\label{L[W]}
\end{equation}
where $W_m$ is the $m$-th harmonics of the field $W(\phi)$.
Note that in Eq.~(\ref{L[W]}) only even harmonics enter; odd harmonics,
as well as the ``longitudinal" modes,
have a larger gap of the order of $\omega_0/\ell$ and are excluded
from the sigma-model action.

Eq.~(\ref{L[W]}) sets a characteristic scale $L$
for the angular variations of matrices $U(\phi)$:
\begin{equation}
\label{L}
  1/L= g \ln g.
\end{equation}
This scale should be larger than the scattering length $\ell$.
Only in this case
one can restrict oneself by the lowest term of the gradient expansion in
powers of $\partial U/\partial\phi$ that has been used in deriving
Eq.~(\ref{sigma-model}). Comparing to $\ell=[g\ln(N/g)]^{-1}$ we see
that the parameter of the gradient expansion, $\ell/L=\ln g/\ln(N/g)$,
is small if the condition (\ref{condition}) is fulfilled.

The length $L$ determines the angular size of the elementary propagator
corresponding to the sigma-model (\ref{sigma-model}). In this respect
it is analogous to the phase-breaking length for Cooperons (or the system
size) in the usual
weak-localization problem. The fact that $\ell/L \ll 1$ in our problem
tells us that the problem is essentially not ballistic, though it is
not diffusive either, since $\ell_{tr}/L=\ln g\gg 1$.

An important property of the action (\ref{sigma-model}) is that it
takes a universal form if $U$ is independent of $\phi$.
It is clear from Eq.~(\ref{L[W]}) that at low energies
 the main contribution
to the functional integral comes from the zero harmonics of $Q(\phi)$,
i.e.\
the problem reduces to the zero-dimensional supermatrix $\sigma$-model.
The uniform supermatrix $Q$ is parametrized by 2 real variables (one of which
appears to be cyclic) and 2 Grassmann variables, so the final expression for
the average DoS is
\begin{equation}
  \langle\rho(\epsilon)\rangle = \frac{1}{4\tilde{\omega}_0} \Re
  \int_0^\pi d\theta \int d\eta d\zeta
  \frac{\sin\theta}{1-\cos\theta}
  \bigl[ (1+\cos\theta) + 2\eta\zeta (1-\cos\theta) \bigr]
  e^{\pi i \frac{\epsilon}{\tilde{\omega}_0} (1-\cos\theta)} =
  \frac{1}{\tilde{\omega}_0} \left(
    1 - \frac{\sin(2\pi\epsilon/\tilde{\omega}_0)}
      {2\pi\epsilon/\tilde{\omega}_0}
  \right) .
\label{rho}
\end{equation}
The functional form of this result coincides with the result of the AZ
phenomenological approach\cite{AZ}. However, it is expressed via the
renormalized mean level spacing
$\tilde{\omega}_0 = \omega_0 \left( 1 + \frac{1}{2\ln g} \right) .
\label{tilde}$
The renormalization is due to the
contribution of
higher $W_{m\geq 2}$ modes which lead\cite{full} to the
decrease of the DoS in the energy range
$\epsilon \leq \omega_0/L = g\omega_0\ln g$
by the relative amount of $\delta \omega_{0}/\omega_{0}=1/(2\ln g) \ll 1$.
At higher energies, the correction decreases as
$\delta \omega_{0}/\omega_{0}\propto (g\omega_0\ln g)^2/\epsilon^2$.
This correction can be found using a general approach \cite{KravMir},
in which the perturbative treatment of the non-zero modes leads to the
``induced" terms in the 0D action. It is given by the {\it single-cooperon}
diagram which is absent in usual normal-metal
problems \cite{efetov,KravMir}; from the formal point of view, the
difference stems from the absence of the BB block in the parametrization
(\ref{W-param}).
A usual \cite{KravMir} {\it two-cooperon} diagram leads to the ``induced"
term $\propto (\epsilon/\omega_0)^2 (g\ln g)^{-1}$
in the effective action of the above 0D $\sigma$-model
(cf.~\cite{AST} where similar result is mentioned).
The possibility to neglect this term determines the upper
limit of energies where purely 0D description is valid:
\begin{equation}
  \epsilon \leq \omega_{Th} = \omega_0\sqrt{g \ln g}.
\end{equation}

To conclude, we have derived microscopically the supersymmetric field
theory for the statistics of the localized electron levels inside the vortex in a moderately clean superconductor.
Our supermatrix $\sigma$-model, Eq.~(\ref{sigma-model}) was {\it derived}
in the main order in the quasiclassical parameter $1/\ln\Delta\tau$.
Previously proposed random-matrix approach \cite{AZ} is shown to be
valid in the low energy range
 $\epsilon \leq \omega_{Th} = [(\omega_0/\tau)\ln(1/\omega_0\tau)]^{1/2}$
where zero-dimensional $\sigma$-model is applicable.
Mixing between zero- and higher modes leads to the decrease
of the DoS by the relative amount of $[2\ln(1/\omega_0\tau)]^{-1}$
at the energies $\epsilon \leq \tau^{-1}\ln(1/\omega_0\tau)$.

Useful discussions with Ya.~M.~Blanter, K.~B.~Efetov,
V.~I.~Fal'ko, Yan V.~Fyodorov, N.~B.~Kopnin,
A.~I.~Larkin, V.~V.~Lebedev, A.~D.~Mirlin, Yu.~V.~Nazarov,  G.~E.~Volovik
are gratefully acknowledged.
This research was supported by the collaboration grant \# 7SUP J048531
from the Swiss NSF, INTAS-RFBR grant \# 95-0302,
RFBR grant \# 98-02-19252, Program ``Statistical Physics" of the Russian
Ministry of Science, DGA grant \# 94-1189 (M.V.F.).
The support from RFBR grant \# 96-02-17133, INTAS-RFBR grant \# 95-0675
and from the U.S. Civilian Research and Development Foundation (CRDF)
under Award \# RP1-209 is gratefully acknowledged (V.E.K.).
M.A.S. acknowledges that this material is based upon
work supported by U.S. Civilian Research and Development
Foundation (CRDF) under Award \# RP1-273.


\end{document}